\documentclass[12pt]{iopart}
\eqnobysec
\usepackage{amssymb}
\usepackage{epsf}
\def\keywords#1{\vspace{10pt}
     \begin{indented}
     \item[]\rm Keywords: #1\par
     \end{indented}}

\def\d{\mathrm d}
\def\exp{\mathrm {exp}}

\def\be{\begin{equation}}
\def\ee{\end{equation}}
\def\bea{\begin{eqnarray}}
\def\eea{\end{eqnarray}}

\begin{document}
\jl{1}

\title[Anomalous diffusion]{Anomalous diffusion in a space- and time-dependent energy landscape}

\author{Lo\"\i c Turban}

\address{Groupe de Physique Statistique, D\'epartement Physique de la Mati\`ere et des Mat\'eriaux,
Institut Jean Lamour\footnote[1]{Laboratoire associ\'e au CNRS UMR 7198.}, CNRS---Nancy Universit\'e---UPV Metz,\\
BP 70239, F-54506 Vand\oe uvre l\`es Nancy Cedex, France}
\ead{loic.turban@ijl.nancy-universite.fr}

\begin{abstract}
We study the influence on diffusion in one dimension of a potential energy perturbation varying as a power in space and time. We concentrate on the case of a parabolic perturbation in space decaying as $t^{-\omega}$ which shows a rich variety of scaling behaviours. When $\omega=1$, the perturbation is truly marginal and leads to anomalous (super)diffusion with a dynamical exponent varying continuously with the perturbation amplitude below some negative threshold value. For slower decay, $\omega<1$, the perturbation becomes relevant and the system is either subdiffusive for an attractive potential or displays a stretched-exponential behaviour for a repulsive one. Exact results are obtained for the mean value and the variance of the position as well as for the surviving probability.
\end{abstract}

\submitto{J. Stat. Mech.}


\section{Introduction} 
Let $X(t)$ denote the position of a diffusing particle at time $t$, anomalous diffusion is characterized by a deviation of the variance $\langle\Delta X^2(t)\rangle$ from the linear time dependence which follows from the central limit theorem~\cite{feller68}. This deviation is generally associated with a change in the Hurst exponent $H$ defined through:
\be 
\langle\Delta X^2(t)\rangle=\langle X^2(t)\rangle-\langle X(t)\rangle^2\sim t^{2H}\,.
\label{1-1}
\ee
Normal diffusion with $H=1/2$ separates the regime of subdiffusion for $0<H<1/2$ from the regime of superdiffusion for $H>1/2$. Note that the limiting value $H=0$ can be associated with a logarithmic subdiffusive behaviour
\be 
\langle\Delta X^2(t)\rangle\sim\ln^\mu t\,.
\label{1-2}
\ee

Anomalous diffusion is observed in various situations\footnote{For a long list of experimental observations of anomalous diffusion see~\cite{metzler00}} among which one may cite turbulent 
diffusion~\cite{richardson26}, transport in amorphous semiconductors~\cite{scher75,pfister78}, diffusion in actin networks~\cite{wong04} and living cells~\cite{golding06,szymanski09}, more generally transport in disordered media as well as on fractals~\cite{havlin87}--\cite{hughes95a}.

Different models have been proposed to explain anomalous diffusion. Subdiffusive behaviour is obtained with the continuous-time random walk~\cite{montroll65}--\cite{hughes95b} in which the random walker jumps between lattice sites after waiting for some random time $\tau$. When the waiting time has a broad distribution (``fat tail"), behaving as $\psi(\tau)\sim\tau^{-1-2H}$ with $0<H\leq 1/2$, then the variance behaves as in equation \eref{1-1}, with a logarithmic correction to the normal behaviour when $H=1/2$ \cite{montroll73}--\cite{shlesinger74}.

The problem of anomalous diffusion has been modelled using fractional dynamics, a very active field in the last 
years~\cite{metzler00,metzler04}. In this approach, fractional derivatives replace ordinary ones in standard diffusion equations. One may also mention generalized Langevin equations which display anomalous diffusion when
the noise correlation function and the dissipative kernel have long-time tails~\cite{porra96}.

Models of random walks with memory may lead to subdiffusive, normal or superdiffusive behaviour, depending on the value of a memory parameter~\cite{schutz04,kumar10}. Superdiffusive behaviour may also result from the presence of long-range step-step correlations~(see \cite{bouchaud90} p 148).

Exact results have been obtained for diffusion with quenched randomness in one dimension (Sinai model) where asymmetric random transition rates lead to the logarithmic behaviour of equation \eref{1-2} with $\mu=4$~\cite{sinai82}. When the transition rates follow some self-similar aperiodic sequence, the diffusion is normal as long as the fluctuations of the environment, characterized by a wandering exponent $\Omega$~\cite{luck93}, are bounded ($\Omega<0$). When the fluctuations are unbounded ($\Omega>0$),  logarithmic diffusion is recovered with $\mu=2/\Omega$\footnote{Note that $\Omega=1/2$ for a random sequence, which leads to Sinai's result $\mu=4$ as expected.}. In the marginal case $\Omega=0$  diffusion is governed by equation \eref{1-1}, with $H$ varying with the amplitude of the aperiodic perturbation~\cite{igloi99}. 

The same scenario occurs in the field of critical phenomena for the Hilhorst-van Leeuwen (HvL) model~\cite{hilhorst81}.
A perturbation decaying as a power of the distance from a free surface  may be irrelevant, marginal or relevant depending on the value of the decay exponent, which plays the same role as 
$\Omega$~\cite{hilhorst81}--\cite{igloi93}.
It leads to continuously varying surface exponents in the marginal case and to a stretched-exponential surface critical behaviour when the perturbation is relevant. Alternatively, the perturbation may decay in time with similar effects, this has been considered in the case of reaction-diffusion processes with a time-dependent reaction rate~\cite{dorogovtsev01,turban04}.

In the present work we introduce a model for anomalous diffusion, where  diffusion takes place in a space- and time-dependent potential energy landscape. The perturbation leads to a rich variety of dynamical behaviours, depending on the values of the exponents which govern its space and time dependence. Truly marginal behaviour is obtained when the potential is parabolic in space and decays as $1/t$ in time. Then, like the surface magnetic and thermal critical exponents of the HvL model~\cite{peschel84}--\cite{igloi93}, the dynamical exponent varies continuously with the perturbation amplitude below some threshold value at which the behaviour is logarithmic.
 
The paper is organized as follows. The model and the associated Fokker-Planck equation are introduced in section~2. Some scaling considerations are given in section~3. Exact results about anomalous diffusion in an energy landscape which is parabolic in space are presented in section~4. We solve the diffusion equation in section~4.1. We study the mean value and variance of $X(t)$ in section~4.2 and the surviving probability in section~4.3. This is followed by a discussion in section~5. Some mathematical details are relegated to the appendices.

\section{Fokker-Planck equation}

We consider a particle diffusing in one space dimension in a space- and time-dependent  energy landscape
\be
U(X,t)=A\,\frac{X^p}{p}\,t^{-\omega}\,,\qquad -\infty<X<+\infty\,,\qquad t>0\,,
\label{2-1}
\ee
with $p\in\mathbb{Z}^*$ and $\omega\in\mathbb{R}$.
The associated force is
\be
F_X=-\frac{\partial U}{\partial X}=-A\,X^{p-1}\,t^{-\omega}
\label{2-2}
\ee
and the drift velocity is given by the Nernst-Einstein relation
\be
V_X=\frac{DF_X}{k_{\rm B}T}=-C\,X^{p-1}\,t^{-\omega}\,,\qquad C=\frac{DA}{k_{\rm B}T}\,,
\label{2-3}
\ee
where $D$ is the diffusion constant, $k_{\rm B}$ is the Boltzmann constant and $T$ the temperature.
Let $P(X,t)\,\d X$ be the probability to find the particle between $X$ and $X+\d X$ at time $t$, the probability current is the sum of the diffusion and drift contributions
\be 
J_X=-D\,\frac{\partial P}{\partial X}+PV_X
\label{2-4}
\ee
and the conservation of probability leads to:
\be
\frac{\partial P}{\partial t}=-\frac{\partial J_X}{\partial X}
=D\,\frac{\partial^2 P}{\partial X^2}-V_X\,\frac{\partial P}{\partial X}
-P\,\frac{\partial V_X}{\partial X}\,.
\label{2-5}
\ee
Combining equations \eref{2-3} and \eref{2-5} one finally obtains the Fokker-Planck equation
\be
\frac{\partial P}{\partial t}=D\,\frac{\partial^2 P}{\partial X^2}+C\,X^{p-1}\,t^{-\omega}\,\frac{\partial P}{\partial X}
+(p-1)CX^{p-2}\,t^{-\omega}\,P\,.
\label{2-6}
\ee

\section{Scaling considerations}
The diffusion is normal in the absence of a drift, i.e., when $C=0$. Then, under a change of the length scale by a factor $b>1$, one has
\be
X'=\frac{X}{b}\,,\qquad t'=\frac{t}{b^z}\,,
\label{3-1}
\ee
and the dynamical exponent $z=1/H=2$ at the unperturbed fixed point. Note that $t$ scales like $X^2$ so that $D$ is dimensionless.

It follows from a dimensional analysis that $V_X/X$ scales like $1/t$. Thus, according to \eref{2-3}
\be
C'\,(X')^{p-2}(t')^{-\omega}=b^z\,C\,X^{p-2}t^{-\omega}\,.
\label{3-2}
\ee
Making use of \eref{3-1}, one obtains the scaling behaviour of the perturbation amplitude $C$ which transforms as\footnote{This scaling behaviour remains valid for non-zero real values of $p$ provided $X$ is replaced by its absolute value in \eref{2-1}.}:
\be
C'=b^{z(1-\omega)+p-2}\,C\,,\qquad z=2\,.
\label{3-3}
\ee
When $\omega>p/2$ the perturbation is irrelevant since $C$ has a negative scaling dimension and decreases under rescaling. Then one expects normal diffusion, at least for small $C$ values.

When $\omega=p/2$ the scaling dimension of $C$ vanishes and the perturbation is marginal. The diffusion should then be governed by a fixed line parameterized by $C$ and leading to continuously varying exponents if the perturbation is truly marginal.

When $\omega<p/2$ the perturbation decays sufficiently slowly in time to become relevant. $C$ increases under rescaling and the normal diffusion fixed point is unstable. The perturbation then drives the system to a new fixed point with a different scaling behaviour.

Let $\phi(X,t,C)$ be some physical quantity with scaling dimension $x_\phi$. According to equations \eref{3-1} and \eref{3-3}, $\phi$ transforms as
\be
\phi'=\phi(X',t',C')=\phi\left(\frac{X}{b},\frac{t}{b^2},b^{\,p-2\omega}C\right)
=b^{\,x_\phi}\phi(X,t,C)\,,
\label{3-4}
\ee
at the normal diffusion fixed point. With $b=|C|^{-1/(p-2\omega)}$ one obtains
\be
\phi(X,t,C)=X_C^{-x_\phi}\phi\left(\frac{X}{X_C},\frac{t}{t_C},{\mathrm {sgn}}(C)\right)
\label{3-5}
\ee
where
\be
X_C=|C|^{-1/(p-2\omega)}\,,\qquad t_C=X_C^2=|C|^{-2/(p-2\omega)}\,
\label{3-6}
\ee
are characteristic length and time introduced by the perturbation when $\omega\neq p/2$. $X_C$ and $t_C$ diverge for a relevant perturbation and vanish for an irrelevant perturbation when $C\to0$.

\section{Parabolic energy landscape in space}

\subsection{Solution of the Fokker-Planck equation}
When $p=2$ the Fokker-Planck equation \eref{2-6} becomes
\be
\frac{\partial P}{\partial t}=D\,\frac{\partial^2 P}{\partial X^2}+C\,X\,t^{-\omega}\,\frac{\partial P}{\partial X}
+C\,t^{-\omega}\,P\,.
\label{4-1}
\ee
Let us introduce the new variables
\bea
&\xi =\xi (X,t)=X\e^{\,f(t)}\,,\qquad \tau=\tau(t)=\int_{t_0}^t\e^{2f(t')}\,\d t'\,,\nonumber\\
&f(t)=C\int_{t_0}^t(t')^{-\omega}\,\d t'=C\,\frac{t^{1-\omega}-t_0^{1-\omega}}{1-\omega},
\label{4-2}
\eea
and the change of function
\be
P(X,t)=\e^{f(t)}\Pi(\xi,\tau)\,,
\label{4-3}
\ee
so that $\Pi(\xi,\tau)$ is a probability density for the new variables.
One has:
\bea
\frac{\partial P}{\partial t}&=C\,t^{-\omega}\,\e^{f(t)}\,\Pi(\xi,\tau)
+C\,X\,t^{-\omega}\,\e^{2f(t)}\,\frac{\partial \Pi}{\partial \xi}
+\e^{3f(t)}\,\frac{\partial \Pi}{\partial \tau}\,,\nonumber\\
\frac{\partial P}{\partial X}&=\e^{2f(t)}\,\frac{\partial \Pi}{\partial \xi}\,,
\qquad \frac{\partial^2 P}{\partial X^2}=\e^{3f(t)}\,\frac{\partial^2 \Pi}{\partial \xi^2}\,.
\label{4-4}
\eea
Collecting these expressions in equation \eref{4-1}, one obtains the ordinary diffusion equation
\be
\frac{\partial \Pi}{\partial \tau}=D\,\frac{\partial^2 \Pi}{\partial \xi^2}\,.
\label{4-5}
\ee
For a particle starting from $\xi=X_0$ at $\tau=0$ (i.e. $X=X_0$ at $t=t_0$ for the original variables, see \eref{4-2}), the solution of equation \eref{4-5} is the Gaussian
\be
\Pi(\xi,\tau)=\frac{1}{\sqrt{4\pi D\tau}}\,\exp\left(-\frac{(\xi-X_0)^2}{4D\tau}\right)\,.
\label{4-6}
\ee
Coming back to the original variables leads to 
\be
P(X,t)=\frac{1}{\sqrt{4\pi DF(t)}}\,\exp\!\left(-\frac{(X\!-\!X_0\e^{-f(t)})^2}{4DF(t)}\right),
\label{4-7} 
\ee
where
\be
F(t)=\e^{-2f(t)}\tau(t)=\!\int_{t_0}^t\!\exp\!\left(\!2C\,\frac{(t')^{1-\omega}\!-\!t^{1-\omega}}{1-\omega}\right)\d t'\,.
\label{4-8}
\ee
This is the solution of equation \eref{4-1} for $t\geq t_0$ corresponding to the initial condition $P(X,t_0)=\delta(X-X_0)$ since $f(t)$ and $F(t)$ both vanish when $t\to t_0$. For other initial conditions at $t=t_0$ the solution can be deduced from \eref{4-7} through convolution.

\subsection{Mean value and variance of the position}
The mean value of the position at time $t$, $\langle X(t)\rangle$, and the variance $\langle\Delta X^2(t)\rangle$ follow from equation \eref{4-7} and, according to~\eref{4-2}, read:
\be\fl
\langle X(t)\rangle=X_0\,\e^{-f(t)}=X_0\,\exp\left(-C\,\frac{t^{1-\omega}-t_0^{1-\omega}} {1-\omega}\right)\,,\qquad
\langle\Delta X^2(t)\rangle=2DF(t)\,.
\label{4-9}
\ee
Let us now examine how they behave when $\omega$ is varied.

\subsubsection{Irrelevant perturbation, $\omega>1$.}
The mean position has the following asymptotic behaviour when $t\gg t_0$:
\be\fl
\langle X(t)\rangle\simeq X_\infty\!\left[1+\mathrm{sgn}(C)\frac{(t_C/t)^{\omega-1}}{\omega-1}\right]\,,\quad
X_\infty=X_0\,\exp\!\left[-\mathrm{sgn}(C)\frac{(t_C/t_0)^{\omega-1}}{\omega-1}\right]\,.
\label{4-10}
\ee
The limiting value $X_\infty$ is smaller (larger) than $X_0$ and is approached from above (below) when $C$ is positive (negative). 

Through the change of variables $u=t'/t$, $F(t)$ in \eref{4-8} can be rewritten as:
\be
F(t)=t\,\exp\left(2C\frac{t^{1-\omega}}{\omega-1}\right)\int_{t_0/t}^1\exp\left(2Ct^{1-\omega}\frac{u^{1-\omega}}{1-\omega}\right)\d u\,.
\label{4-11}
\ee
Expanding the exponential and integrating term by term, one obtains
\bea
\fl F(t)=t\,\exp\left(2C\frac{t^{1-\omega}}{\omega-1}\right)\sum_{k=0}^\infty\frac{1}{k!}
\left(2C\frac{t^{1-\omega}}{1-\omega}\right)^k\int_{t_0/t}^1u^{k(1-\omega)}\d u\,,\nonumber\\
\fl\ \ \ \ \ \ =t\,\exp\left(2C\frac{t^{1-\omega}}{\omega-1}\right)\sum_{k=0}^\infty\frac{2^k}{k!(1-\omega)^k}
\frac{(Ct^{1-\omega})^k-(t_0/t)(Ct_0^{1-\omega})^k}{1+k(1-\omega)}\,,\nonumber\\
\fl\ \ \ \ \ \ =t\,\exp\left(2C\frac{t^{1-\omega}}{\omega-1}\right)\sum_{k=0}^\infty\frac{2^k}{k!(1-\omega)^k}
\frac{(t_C/t)^{k(\omega-1)}-(t_0/t)(t_C/t_0)^{k(\omega-1)}}{1+k(1-\omega)}\,,\nonumber\\
\fl\ \ \ \ \ \ =t\,\left\{1+\mathrm{O}\left[(t_C/t)^{\omega-1}\right]+\mathrm{O}(t_0/t)\right\}\,,
\label{4-12}
\eea
and $F(t)$ behaves as:
\be
F(t)=\left\{ \begin{array}{ll}
t\left\{1+\mathrm{O}[(t_C/t)^{\omega-1}]\right\}& \mbox{when}\ 1<\omega\leq2\\ \ms
t\left[1+\mathrm{O}(t_0/t)\right]& \mbox{when}\ \omega>2
\end{array}\right.\,.
\label{4-13}
\ee
Thus, asymptotically we obtain the normal diffusion behaviour
\be
\langle\Delta X^2(t)\rangle\simeq 2Dt=2D\,X_C^2\frac{t}{t_c}\,,\qquad t\gg t_0,\ t_C\,,
\label{4-14}
\ee
as expected in the presence of an irrelevant perturbation.

\begin{figure} [tbh]
\epsfxsize=10cm
\hskip 16mm
\mbox{\epsfbox{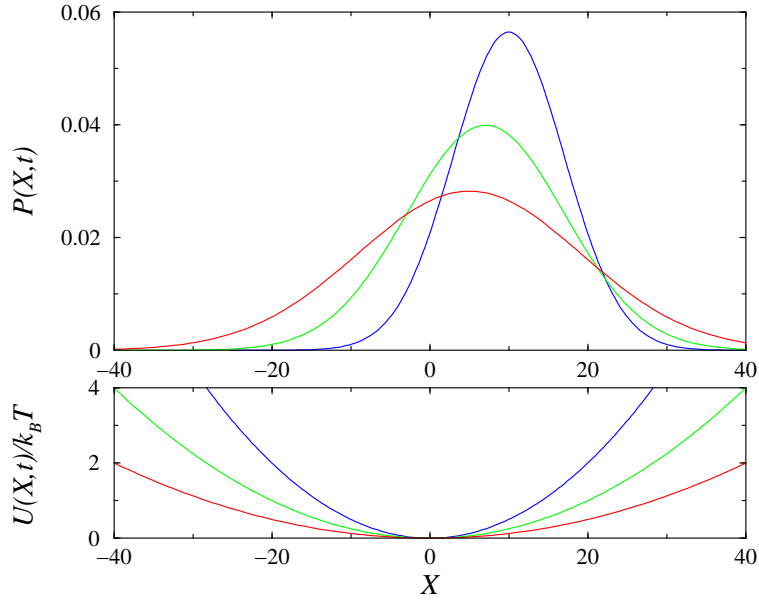}}
\vskip -.2cm
\caption{Anomalous diffusion from the initial position $X_0=100$ at $t_0=1$ in a marginal attractive parabolic potential decaying as $t^{-1}$. The probability density (top) and the reduced potential energy (bottom) are shown for $C=1/2$ and $D=1/2$ at times $t=100$ (blue), 200 (green), 400 (red). The variance has a normal linear time dependence whereas the mean position decays as $1/\sqrt{t}$}
\label{fig1-anom-dif}  \vskip 0cm
\end{figure}

\begin{figure} [tbh]
\epsfxsize=10cm
\hskip 16mm
\mbox{\epsfbox{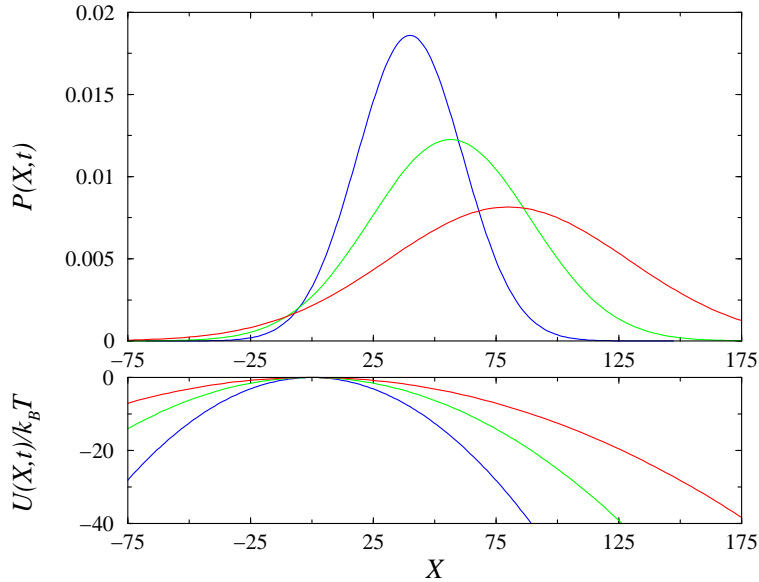}}
\vskip -.2cm
\caption{Anomalous diffusion from the initial position $X_0=4$ at $t_0=1$ in a marginal repulsive parabolic potential decaying as $t^{-1}$. The probability density (top) and the reduced potential energy (bottom) are shown for the threshold value $C=-1/2$ and $D=1/2$  at time $t=100$ (blue), 200 (green), 400 (red). Here the variance has a logarithmic correction and grows as $t\ln t$ whereas the mean position increases as $4\sqrt{t}$.}
\label{fig2-anom-dif}  \vskip 0cm
\end{figure}

\begin{figure} [tbh]
\epsfxsize=10cm
\hskip 16mm
\mbox{\epsfbox{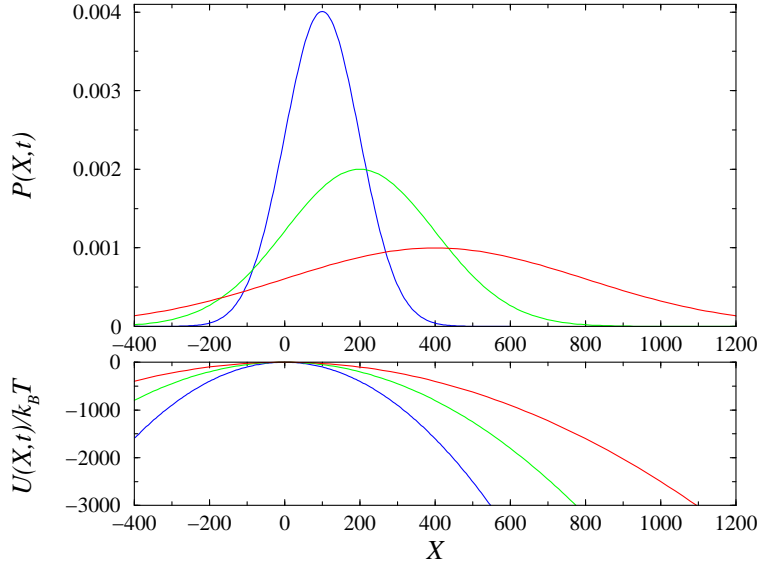}}
\vskip -.2cm
\caption{Anomalous diffusion from the initial position $X_0=1$ at $t_0=1$ in a marginal repulsive parabolic potential decaying as $t^{-1}$. The probability density (top) and the reduced potential energy (bottom) are shown for $C=-1$ and $D=1/2$ at times $t=100$ (blue), 200 (green), 400 (red). For this value of $C<-1/2$ the mean position increases as $t$ and the variance scales like $\langle X(t)\rangle^2\propto t^2$.}
\label{fig3-anom-dif}  \vskip 0cm
\end{figure}

\subsubsection{Marginal perturbation, $\omega=1$.}

When $\omega\to1$, $f(t)\to C\ln(t/t_0)$ so that the integral giving $F(t)$ in \eref{4-8} can be evaluated exactly:
\be
\lim_{\omega\to1}F(t)\!=t^{-2C}\!\!\!\int_{t_0}^t (t')^{2C}\,\d t'=\frac{t}{2C\!+\!1}\left[1\!-\!\left({t_0}/{t}\right)^{2C+1}\right]\,.
\label{4-15}
\ee
Making use of these results in \eref{4-9}, one obtains:
\bea
\fl\langle X(t)\rangle=X_0\left({t}/{t_0}\right)^{-C}\,,\nonumber\\ \ms
\fl\langle\Delta X^2(t)\rangle=\left\{ \begin{array}{ll}
\frac{2Dt}{2C+1}\left[1-\left({t_0}/{t}\right)^{2C+1}\right]& \mbox{when}\ C>-1/2\\ \ms
2Dt\,\ln(t/t_0)& \mbox{when}\ C=-1/2\\ \ms
\frac{2Dt_0}{2|C|-1}\left({t}/{t_0}\right)^{2|C|}\left[1-\left(t_0/t\right)^{2|C|-1}\right]& \mbox{when}\ C<-1/2
\end{array}\right.
\label{4-16}
\eea
The mean position is algebraic, with a continuously varying exponent for any non-zero value of $C$. It decreases to zero when the potential is attractive ($C>0$) and grows to infinity when the potential is repulsive ($C<0$). When $C>-1/2$, the leading behaviour of the variance is normal with a dynamical exponent $z=2$ ($\langle\Delta X^2(t)\rangle\sim t^{2/z}$) whereas the amplitude is $C$-dependent. The marginal behaviour shows up in the correction-to-scaling. A logarithmic correction to the normal diffusion appears at $C=-1/2$. 
Below this threshold value the dynamical exponent is continuously varying, as expected in the presence of a truly marginal perturbation. 
Since $z(C)=1/|C|<2$, a superdiffusive behaviour is obtained when $C<-1/2$.

The time evolution of the marginal probability density and the corresponding reduced potential energy are shown for $C=1/2$ in figure 1, at the threshold value $C=-1/2$ in figure 2 and below the threshold, for $C=-1$, in figure 3. Note the change in the length scales of the three figures.

\subsubsection{Relevant perturbation, $0\leq\omega<1$.}
 
According to \eref{4-9} the mean position behaves as:
\be
\langle X(t)\rangle=X_0\,\exp\left[-\mathrm{sgn}(C)\frac{(t/t_C)^{1-\omega}-(t_0/t_C)^{1-\omega}}{1-\omega}\right]\,.
\label{4-17}
\ee
Thus the mean position vanishes (diverges) as a stretched-exponential function of the time when the potential is attractive (repulsive).

When $C>0$ the main contribution to $F(t)$ in \eref{4-8} comes from the vicinity of $t'=t$. Expanding $f(t')$ near $t$ to second order in $u=t'-t$, one obtains
\bea
\fl F(t)=\int_{t_0-t}^0\e^{2C\,t^{-\omega}u}(1-\omega C\,t^{-\omega-1}u^2+\cdots)\,\d u
=\frac{1-\e^{-2C\,t^{-\omega}(t-t_0)}}{2C\,t^{-\omega}}-\frac{\omega t^{2\omega-1}}{8C^2}\nonumber\\
\fl\ \ \ \ \ \ \ \ \ \ \ \times\left\{2-\e^{-2C\,t^{-\omega}(t-t_0)}\left[4C^2t^{-2\omega}(t-t_0)^2
+4Ct^{-\omega}(t-t_0)+2\right]\right\}+\cdots\nonumber\\
\fl\ \ \ \ \ \ =\frac{t^\omega}{2C}\!\left\{1\!+\mathrm{O}\left[(t_C/t)^{1-\omega}\right]\right\}\,,
\label{4-18}
\eea
when $t\gg t_0$ and $t_C$, in agreement with the exact results of appendix A in the same limit. Thus the variance is given by:
\be
\langle\Delta X^2(t)\rangle\simeq\frac{D}{C}\,t^\omega\simeq D X_C^2\!\left(\frac{t}{t_C}\right)^\omega\,,\qquad t\gg t_0,\ t_C\,.
\label{4-19}
\ee

When $C<0$ we can split the integral giving $\tau(t)$ in two parts such that
\be\fl
\tau(t)=\tau(\infty)-\delta\tau(t)\,,\quad \tau_\infty=\lim_{t\to\infty}\tau(t)=I(t_1=t_0)\,,\quad \delta\tau(t)=I(t_1=t)\,,
\label{4-20}
\ee
with, according to~\eref{4-2}:
\be
I(t_1)=\int_{t_1}^\infty\!\!\exp\left[-2|C|\,\frac{(t')^{1-\omega}-t_0^{1-\omega}} {1-\omega}\right]\,\d t'\,.
\label{4-21}
\ee
This integral is studied in appendix B.

When $t\gg t_C$, $\delta\tau(t)$ is given by \eref{B-5}:
\be
\delta\tau(t)\sim\frac{t^\omega}{2|C|}\exp\left(-2|C|\,\frac{t^{1-\omega}-t_0^{1-\omega}} {1-\omega}\right)\left\{1+\mathrm{O}\left[(t_C/t)^{1-\omega}\right]\right\}\,.
\label{4-22}
\ee

When $t>t_0\gg t_C$, $\tau_\infty$ is given by~\eref{B-5}, too, and reads:
\be
\tau_\infty=\frac{t_0^\omega}{2|C|}\left\{1+\mathrm{O}\left[(t_C/t_0)^{1-\omega}\right]\right\}\,.
\label{4-23}
\ee
Thus, using \eref{4-8}, we obtain:
\be
F(t)\sim\frac{t_0^\omega}{2|C|}\,\exp\left(2|C|\frac{t^{1-\omega}-t_0^{1-\omega}}{1-\omega}\right)-\frac{t^\omega}{2|C|}\,.
\label{4-24}
\ee
This expression is valid for large values of $|C|$ such that $t_C\ll t$ and $t_C\ll t_0$. It agrees with the exact results in appendix A in the same limit.

When $t\gg t_C\gg t_0$, $\tau_\infty$ is given by~\eref{B-7}
\be
\tau_\infty=a_\omega\,\frac{t_C^\omega}{2|C|}\left[1+\mathrm{O}(t_0/t_C)\right]\,,\qquad 
a_\omega=\left(\frac{1-\omega}{2}\right)^{\frac{\omega}{1-\omega}}\Gamma\!\left(\frac{1}{1-\omega}\right)\,,
\label{4-25}
\ee
so that:  
\be
F(t)\sim a_\omega\,\frac{t_C^\omega}{2|C|}
\exp\!\left(\frac{2|C|\,t^{1-\omega}} {1-\omega}\right)-\frac{t^\omega}{2|C|}\,,
\label{4-26}
\ee
once more in agreement with the results of appendix A.

Collecting these results in \eref{4-9} gives the long-time behaviour of the variance
\be
\langle\Delta X^2(t)\rangle\sim
\frac{D}{|C|}\left[t_0^\omega\,\exp\left(2|C|\frac{t^{1-\omega}-t_0^{1-\omega}}{1-\omega}\right)-t^\omega\right]\,\quad t>t_0\gg t_C\,,
\label{4-27}
\ee
and
\be
\langle\Delta X^2(t)\rangle\!\sim\!\frac{D}{|C|}\!\left[a_\omega\,t_C^\omega
\,\exp\!\left(\frac{2|C|\,t^{1-\omega}} {1-\omega}\right)-t^\omega\right]\,,\  t\gg t_C\gg t_0\,.
\label{4-28}
\ee
In terms of scaled variables, we obtain
\be\fl
\frac{\langle\Delta X^2(t)\rangle}{X_C^2}\sim D\left[\left(\frac{t_0}{t_C}\right)^\omega
\!\!\exp\!\left(2\,\frac{(t/t_C)^{1-\omega}\!-\!(t_0/t_C)^{1-\omega}}{1-\omega}\right)\!-\!
\left(\frac{t}{t_C}\right)^\omega\right],\  t>t_0\gg t_C\,,
\label{4-29}
\ee
and
\be
\frac{\langle\Delta X^2(t)\rangle}{X_C^2}\sim\!D\left[ a_\omega
\,\exp\!\left(\frac{2(t/t_C)^{1-\omega}} {1-\omega}\right)-\left(\frac{t}{t_C}\right)^{\omega}\right],\quad  t\gg t_C\!\gg t_0\,.
\label{4-30}
\ee

With an attractive potential ($C>0$), the variance is subdiffusive with an $\omega$-dependent exponent whereas it is superdiffusive with a stretched-exponential behaviour when the potential is repulsive ($C<0$).

\subsection{Surviving probability}

\begin{table}\fl
\caption{\label{t1}Variation with $\omega$ and $C$ of the long-time behaviour ($t\gg t_C\gg t_0$) of the mean position $\langle X(t)\rangle$, the variance $\langle\Delta X^2(t)\rangle$ and the surviving probability ${\cal S}(t)$ for the diffusion in a parabolic potential decaying as $t^{-\omega}$.}
\begin{indented}\item[]
\begin{tabular}{@{}lllll}
\br
$\omega$&$C$ &$\langle X(t)\rangle$&$\langle\Delta X^2(t)\rangle$&${\cal S}(t)$\\
\mr
$>\!1$&Any&$X_0\,\exp\!\left(-\frac{C\,t_0^{1-\omega}}{\omega-1}\right)$&$2Dt$&$\propto t^{-1/2}$\\ \ms
1&$>\!\!-1/2$&$X_0\,(t/t_0)^{-C}$&$\frac{2D}{2C+1}t$&$\propto t^{-C-1/2}$\\ \ms
1&$-1/2$&$X_0\,\sqrt{t/t_0}$&$2Dt\ln(t/t_0)$&$\propto (\ln t)^{-1/2}$\\ \ms
1&$\!\!<-1/2$&$X_0\,(t/t_0)^{|C|}$&$\frac{2D}{2|C|-1}t_0(t/t_0)^{2|C|}$&${\cal S}_\infty+\alpha t^{-2|C|+1}$\\ \ms
$<\!1$&$>0$&$X_0\,\exp\!\left(-C\frac{t^{1-\omega}}{1-\omega}\right)$&$\frac{D}{C}\,t^\omega$&$\propto t^{-\omega/2}\e^{-\kappa t^{1-\omega}}$\\ \ms
$<\!1$&$<0$&$X_0\,\exp\!\left(|C|\frac{t^{1-\omega}}{1-\omega}\right)$&$a_\omega\frac{D}{|C|}\,t_C^\omega\,
\exp\!\left(2|C|\frac{t^{1-\omega}}{1-\omega}\right)$&${\cal S}_\infty+\beta t^\omega\e^{-2\kappa t^{1-\omega}}$\\
\br
\end{tabular}
\end{indented}
\end{table}

In order to study the surviving probability in the half-space $X>0$ of a particle starting from $X_0>0$ at $t_0$ we introduce an absorbing boundary condition at $X=0$, $P(0,t)=0$. The solution of the Fokker-Planck equation~\eref{4-1} with a parabolic potential is obtained using the image method and reads
\be\fl
P(X,t)=\frac{1}{\sqrt{4\pi DF(t)}}\left[\exp\!\left(-\frac{(X\!-\!X_0\,\e^{-f(t)})^2}{4DF(t)}\right)
-\exp\!\left(-\frac{(X\!+\!X_0\,\e^{-f(t)})^2}{4DF(t)}\right)\right]
\label{4-31}
\ee
according to \eref{4-7}. Let ${\cal F}(t)\,\d t$ give the probability that the diffusing particle is absorbed at $X=0$ between $t$ and $t+\d t$, the first-passage probability density is given by~\cite{redner01}
\be\fl
{\cal F}(t)=-J_X(0,t)=D\,\frac{\partial P(X,t)}{\partial X}|_{X=0}
=\frac{X_0\,\e^{2f(t)}}{\sqrt{4\pi D\tau^3(t)}}\,\exp\left(-\frac{X_0^2}{4D\tau(t)}\right)
\label{4-32}
\ee
Note that the drift term does not contribute directly to $J_X(0,t)$ since $V_X(0,t)$ vanishes when $p=2$ (see equation~\eref{2-3}). The surviving probability can be written as:
\be
{\cal S}(t)=\int_0^\infty \!\!P(X,t)\,\d X=1-\int_{t_0}^t\!{\cal F}(t')\,\d t'\,.
\label{4-33}
\ee
In the last integral the change of variable $u=X_0/\sqrt{4D\tau(t')}$ leads to
\be
\int_{t_0}^t\!{\cal F}(t')\,\d t'=\frac{2}{\sqrt{\pi}}\int_{{X_0}/{\sqrt{4D\tau(t)}}}^\infty\e^{-u^2}\d u\,,
\label{4-34}
\ee
so that
\be
{\cal S}(t)=\frac{2}{\sqrt{\pi}}\int_0^{{X_0}/{\sqrt{4D\tau(t)}}}\e^{-u^2}\d u
=\mathrm{erf}\left(\frac{X_0}{\sqrt{4D\tau(t)}}\right)\,,
\label{4-35}
\ee
where $\mathrm{erf}(x)$ is the error function (see \cite{abramowitz65} p~297)  with the series expansion:
\be
\mathrm{erf}(x)=\frac{2}{\sqrt{\pi}}\sum_{n=0}^\infty(-1)^n\frac{x^{2n+1}}{n!(2n+1)}\,.
\label{4-36}
\ee

In the following we apply these results in the different regimes.

\subsubsection{Irrelevant perturbation, $\omega>1$.}

Inserting \eref{4-13} into \eref{4-8}, to leading order, one obtains
\be
\tau(t)\simeq t\,\exp\!\left(\frac{2C\,t_0^{1-\omega}}{\omega-1}\right)\,,
\label{4-37}
\ee
and \eref{4-35} gives:
\be
{\cal S}(t)\simeq\mathrm{erf}\left[\frac{X_0}{\sqrt{4Dt}}\,\exp\!\left(-\frac{C\,t_0^{1-\omega}}{\omega-1}\right)\right]\,.
\label{4-38}
\ee
As expected for an irrelevant perturbation, the time dependence is the same as in the solution of the unperturbed problem~\cite{redner01} which coincides with~\eref{4-38} when $C=0$.
In the limit $t\to\infty$, the surviving probability ${\cal S}_\infty$ vanishes, even in the presence of a repulsive potential.
A $t^{-1/2}$ long-time behaviour is given by the first term in the series expansion~\eref{4-36}:
\be
{\cal S}(t)\simeq\! \frac{1}{\pi D}\frac{X_0}{X_C}\sqrt{\!\frac{t_C}{t}}\,\exp\!\left(-\mathrm{sgn}(C)\frac{(t_C/t_0)^{\omega-1}}{\omega-1}\right)\,,
\qquad t\gg t_0,t_C\,.
\label{4-39}
\ee

\subsubsection{Marginal perturbation $\omega=1$.}

\begin{figure} [tbh]
\epsfxsize=10cm
\hskip 16mm
\mbox{\epsfbox{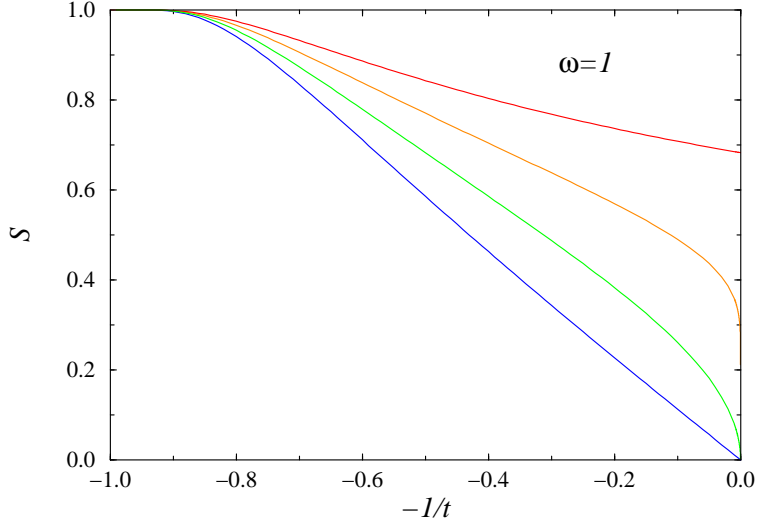}}
\vskip -.2cm
\caption{Time dependence of the surviving probability of a particle starting from $X_0=1$ at $t_0=1$ in the marginal case ($\omega=1$) with $C=1/2$ (blue), $C=0$ (green), $C=-1/2$ (orange) and $C=-1$ (red). ${\cal S}$ decays very slowly as $(\ln t)^{-1/2}$ when $C=-1/2$. The decay is algebraic when $C\neq-1/2$, with a non-vanishing asymptotic value when $C<-1/2$.}
\label{fig4-anom-dif}  \vskip 0cm
\end{figure}

From equations~\eref{4-8} and~\eref{4-15} we deduce:
\be\fl
\tau(t)=t_0\,\frac{(t/t_0)^{2C+1}-1}{2C+1}
\simeq\left\{ \begin{array}{ll}
\frac{t_0}{2C+1}\,(t/t_0)^{2C+1}& \mbox{when}\ C>-1/2\\ \ms
t_0\,\ln(t/t_0)& \mbox{when}\ C=-1/2\\ \ms
\frac{t_0}{2|C|-1}[1-(t_0/t)^{2|C|-1}]& \mbox{when}\ C<-1/2
\end{array}\right.\,.
\label{4-40}
\ee
The surviving probability is given by:
\be
{\cal S}(t)=\left\{ \begin{array}{ll}
\mathrm{erf}\left(\frac{X_0}{\sqrt{4Dt_0}}\sqrt{\frac{2C+1}{(t/t_0)^{2C+1}-1}}\,\right) &\mathrm{when}\ C\neq -1/2\\ \ms
\mathrm{erf}\left(\frac{X_0}{\sqrt{4Dt_0\ln(t/t_0)}}\right) &\mathrm{when}\ C=-1/2
\end{array}\right.\,.
\label{4-41}
\ee
In the long-time limit, $t\gg t_0$, one obtains:
\be\fl
{\cal S}(t)\simeq\left\{ \begin{array}{ll}
X_0\sqrt{(2C+1)/(\pi Dt_0)}\,(t_0/t)^{C+1/2}&\mbox{when}\ C>-1/2\\ \ms
X_0/\sqrt{\pi Dt_0\,\ln(t/t_0)}&\mbox{when}\ C=-1/2\\ \ms
\mathrm{erf}\!\left(X_0\sqrt{(2|C|\!-\!1)/(4 Dt_0)}\left[1\!+\!\frac{1}{2}(t_0/t)^{2|C|-1}\!\right]\right)&\mbox{when}\ C<-1/2
\end{array}\right.\,.
\label{4-42}
\ee
${\cal S}_\infty$ vanishes when $C\geq-1/2$. Only when $C<-1/2$, i.e., when the repulsive force is strong enough, does the surviving probability have a non-vanishing limiting value:
\be
{\cal S}_\infty=\mathrm{erf}\left(X_0\,\sqrt{\frac{2|C|-1}{4 Dt_0}}\right)\,,\qquad C<-1/2\,.
\label{4-43}
\ee
A first-order expansion of the error function gives the deviation from this limiting value:
\be\fl
\delta{\cal S}(t)={\cal S}(t)-{\cal S}_\infty\simeq X_0\,\sqrt{\frac{2|C|\!-\!1}{4\pi Dt_0}}\,\exp\!\left[-\frac{(2|C|\!-\!1)X_0^2}{4Dt_0}\right]\left(\frac{t_0}{t}\right)^{2|C|-1}\!\!\!\!\!\!\!\!,\quad C<-1/2\,.
\label{4-44}
\ee
Thus $\delta{\cal S}(t)$ exhibits an algebraic behaviour with different exponents above ($t^{-C-1/2}$) and below ($t^{-2|C|+1}$) the critical value $C=-1/2$ where the decay is logarithmic [$(\ln t)^{-1/2})$]. 

The time dependence of the surviving probability in the marginal case for different values of $C$ is shown in figure 4.

\subsubsection{Relevant perturbation, $0\leq\omega<1$.}

\begin{figure} [tbh]
\epsfxsize=10cm
\hskip 16mm
\mbox{\epsfbox{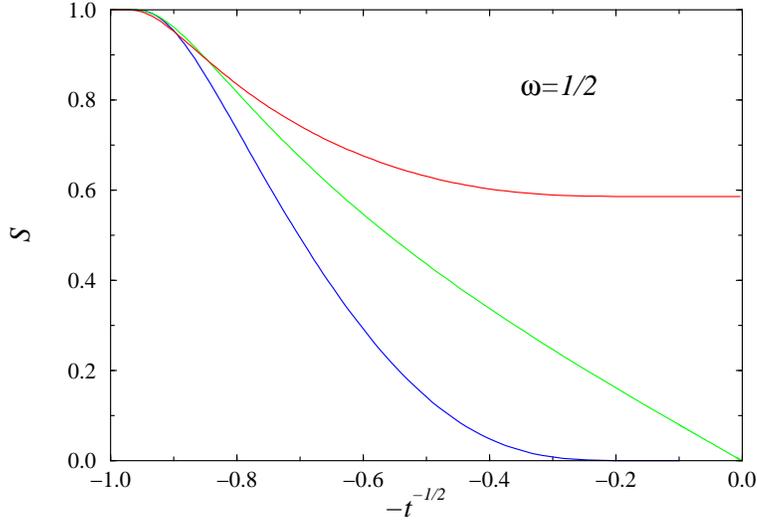}}
\vskip -.2cm
\caption{Time dependence of the surviving probability of a particle starting from $X_0=1$ at $t_0=1$ in the relevant case ($\omega=1/2$) with $C=1$ (blue), $C=0$ (green) and $C=-1/2$ (red). When $C=0$, the diffusion is normal and the surviving probability decays asymptotically as $t^{-1/2}$. There is a stretched-exponential decay when $C\neq0$, with a non-vanishing asymptotic value when $C<0$.}
\label{fig5-anom-dif}  \vskip 0cm
\end{figure}

When $C>0$ and $t\gg t_0$ and $t_C$, equations~\eref{4-8} and~\eref{4-18} lead to:
\be
\tau(t)\simeq \frac{t^\omega}{2C}\,\exp\left(\frac{2C\,t^{1-\omega}}{1-\omega}\right)\,.
\label{4-45}
\ee
The surviving probability behaves as:
\be
{\cal S}(t)\simeq\mathrm{erf}\left[X_0\sqrt{\frac{C}{2Dt^\omega}}\,\exp\left(-\frac{C\,t^{1-\omega}}{1-\omega}\right)\right]\,.
\label{4-46}
\ee
The asymptotic decay to ${\cal S}_\infty=0$ has the following stretched-exponential form:
\be
{\cal S}(t)\simeq\sqrt{\frac{2}{\pi D}}\frac{X_0}{X_C}\left(\frac{t_C}{t}\right)^{\omega/2}\exp\!\left(-\frac{(t/t_C)^{1-\omega}}{1-\omega}\right)\,.
\label{4-47}
\ee

When $C<0$ and $t\gg t_0\gg t_C$ equations~\eref{4-22} and~\eref{4-23} give:
\be
\tau(t)\simeq\frac{t_0^\omega}{2|C|}-\frac{t^\omega}{2|C|}\,\exp\left(-\frac{2|C|t^{1-\omega}}{1-\omega}\right)\,.
\label{4-48}
\ee
Hence, the surviving probability behaves as:
\be
{\cal S}(t)\simeq\mathrm{erf}\left(X_0\sqrt{\frac{|C|}{2Dt_0^\omega}}
\left[1+\frac{1}{2}\left(\frac{t}{t_0}\right)^\omega\exp\left(-\frac{2|C|t^{1-\omega}}{1-\omega}\right)\right]\right)\,.
\label{4-49}
\ee
It follows that:
\be
{\cal S}_\infty\simeq\mathrm{erf}\!
\left(\frac{1}{\sqrt{2 D}}\frac{X_0}{X_C}\left(\frac{t_C}{t_0}\right)^{\omega/2}\right)\,.
\label{4-50}
\ee
A first-order expansion of the error function gives the deviation from ${\cal S}_\infty$, which decays as a stretched exponential:
\be
\fl \delta{\cal S}(t)\simeq\! \frac{1}{\sqrt{2\pi D}}\frac{X_0}{X_C}\!\left(\frac{t_C}{t_0}\right)^{\omega/2}
\!\!\!\!\!\!\exp\!\left[-\frac{1}{2D}\!\left(\frac{X_0}{X_C}\right)^2\!
\left(\frac{t_C}{t_0}\right)^{\omega}\right]
\!\!\left(\frac{t}{t_0}\right)^\omega\!\!\!\exp\!\left(\!\!-\frac{2(t/t_C)^{1-\omega}}{1-\omega}\right).
\label{4-51}
\ee

When $C<0$ and $t\gg t_C\gg t_0$ equations~\eref{4-22} and~\eref{4-25} give:
\be
\tau(t)\simeq a_\omega\frac{t_C^\omega}{2|C|}
-\frac{t^\omega}{2|C|}\,\exp\left(-\frac{2|C|t^{1-\omega}}{1-\omega}\right)\,.
\label{4-52}
\ee
Comparing with equation \eref{4-48}, we note that the only change is the substitution of $a_\omega t_C^\omega$ for $t_0^\omega$.
Hence, the asymptotic surviving probability now reads
\be 
{\cal S}_\infty\simeq\mathrm{erf}\!\left(\frac{1}{\sqrt{2a_\omega D}}\frac{X_0}{X_C}\right)\,.
\label{4-53}
\ee
and the deviation from ${\cal S}_\infty$ displays the same stretched-exponential decay as above:
\be
\fl\delta{\cal S}(t)\simeq\! \frac{1}{\sqrt{2\pi a_\omega^3 D}}\frac{X_0}{X_C}
\,\exp\!\left[-\frac{1}{2a_\omega D}\!\left(\frac{X_0}{X_C}\right)^2\right]
\!\!\left(\frac{t}{t_C}\right)^\omega\!\!\!\exp\!\left(\!\!-\frac{2(t/t_C)^{1-\omega}}{1-\omega}\right).
\label{4-54}
\ee

The time dependence of the surviving probability in the presence of a relevant perturbation ($\omega=1/2$) is shown in figure 5 for different values of $C$.

\section{Discussion}

In this work we have shown that  diffusion may become anomalous under the influence a potential $U(X,t)$ varying as $X^{p}$ in space and decaying as  $t^{-\omega}$ in time. Scaling considerations indicates that the perturbation is relevant (irrelevant) when $\omega<p/2$ ($\omega>p/2$). Then the perturbation introduces a characteristic length $X_C$ and a characteristic time $t_C=X_C^2$, varying as powers of the perturbation amplitude $C$. The scaling functions can be expressed in terms of the scaled variables $X/X_C$ and $t/t_C$. When $\omega=p/2$, $C$ is dimensionless and the perturbation is marginal.

Actually, truly marginal behaviour with $C$-dependent exponents is only obtained for $p=2$. A great variety of scaling behaviours then shows up when $\omega$ and $C$ are varied. Our results are collected in table 1. 

The peculiarity of the parabolic potential appears clearly in equation~\eref{3-3}.  For $p=2$, the perturbation amplitude $C$ scales as 
\be
C'=b^{z(1-\omega)}C\,,
\label{5-1}
\ee
and thus is marginal when $\omega=1$. Since the dynamical exponent enters as a factor of $(1-\omega)$ into the scaling dimension of $C$, the perturbation remains marginal even when $z$ varies with $C$. This is no longer true for any other value of $p$ where a variation of $z$ would immediately lead to a non-vanishing scaling dimension of the perturbation\footnote{In this respect the HvL model is different, since there the marginal behaviour affects the surface magnetic exponents which are continuously varying, whereas it is governed by invariant bulk exponents~\cite{cordery82,burkhardt82a}.}. This may be verified with a linear potential ($p=1$ in equation~\eref{2-6}) where the problem can be solved with the change of variables
\be
\xi=X+C\,\frac{t^{1-\omega}-t_0^{1-\omega}}{1-\omega}\,,\qquad \tau=t-t_0\,,
\label{5-2}
\ee
leading to 
\be
\langle X(t)\rangle=X_0-C\,\frac{t^{1-\omega}-t_0^{1-\omega}}{1-\omega}\,,\qquad \langle\Delta X^2(t)\rangle=2D(t-t_0)\,,
\label{5-3}
\ee
for $t\geq t_0$. Here nothing special occurs at $\omega=1/2$ when the scaling dimension of the perturbation at the unperturbed fixed point vanishes.

With the marginal parabolic potential one may notice  that even if the variance behaves normally when $C>-1/2$ there are traces of the marginal behaviour showing up in the correction-to-scaling of the variance as well as in the mean position which grows or decays as $t^{-C}$, depending on the sign of $C$. The marginal behaviour is apparent, for the variance and for the mean position as well, when $C\leq-1/2$. Actually $\langle\Delta X^2(t)\rangle$ scales like $\langle X(t)\rangle^2$ when $C<-1/2$.

In the relevant situation, $\omega<1$, the mean position always displays a stretched-exponential behaviour. The variance behaves quite differently for attractive and repulsive potentials as in the HvL model \cite{burkhardt84}. In the first case, $C>0$, $\langle\Delta X^2(t)\rangle$ increases as $t^\omega$ whereas the growth is of the stretched-exponential type with the same scaling as $\langle X(t)\rangle^2$ when $C<0$. 

The time evolution of the surviving probability with (i) in the marginal situation, a logarithmic behaviour at $C=-1/2$, an algebraic decay to ${\cal S}_\infty$ with $C$-dependent exponents and a non-vanishing value of ${\cal S}_\infty$ when $C<-1/2$ and (ii) in the relevant situation, a stretched-exponential decay to ${\cal S}_\infty$ and, when $C<0$, a non-vanishing value of ${\cal S}_\infty$, is the same as the 
temperature evolution of the surface magnetization $m_s(T_c-T)$ of the HvL model in the ordered phase~\cite{peschel84,bloete85} with $1/t$ playing the role of $T_c-T$. 

Note that, for the transverse-field Ising model in one dimension, similar correspondences have been noticed before between the finite-size properties of the surviving probability of a random walk in a random or aperiodic environment and the finite-size behaviour of the surface magnetization of the Ising model in the same environment~\cite{igloi98a}--\cite{karevski99}.

Finally let us mention that the diffusion problem can be solved in the same way in higher dimensions for a time-dependent spherically symmetric potential.
 
\appendix
\setcounter{section}{0}

\section{Calculation of $F(t)$ when $\omega=0$ and $1/2$}
\label{appendix-A}

When $\omega=0$ the function $F(t)$ defined in equation~\eref{4-8} yields:
\be
F(t)=\e^{-2Ct}\!\int_{t_0}^t\!\e^{2Ct'}\d t'=\frac{1-\exp\left[-2C(t-t_0)\right]}{2C}\,.
\label{A-1}
\ee
When $\omega=1/2$ we have:
\be
F(t)=\e^{-4C\sqrt{t}}\!\int_{t_0}^t\!\e^{4C\sqrt{t'}}\d t'\,.
\label{A-2}
\ee
Thus, using
\be
\int\e^{a\sqrt{u}}\,\d u=\frac{2}{a^2}\e^{a\sqrt{u}}(a\sqrt{u}-1)\,,
\label{A-3}
\ee
one obtains:
\be
F(t)=\frac{1}{8C^2}\left[(4C\sqrt{t}-1)-\e^{-4C(\sqrt{t}-\sqrt{t_0})}(4C\sqrt{t_0}-1)\right]\,.
\label{A-4}
\ee
The approximate results in equations~\eref{4-18}, \eref{4-24} and~\eref{4-26} are 
all in accordance with these exact results when the appropriate limits are taken.

\section{Calculation of $I(t_1)$}
\label{appendix-B}

According to \eref{4-21} we have:
\be
I(t_1)=\exp\left(\frac{2|C|\,t_0^{1-\omega}} {1-\omega}\right)\int_{t_1}^\infty\!\!\exp\left(-\frac{2|C|\,(t')^{1-\omega}} {1-\omega}\right)\,\d t'\,.
\label{B-1}
\ee
With the change of variable $u(t')=2|C|(t')^{1-\omega}/(1-\omega)$, $I(t_1)$ can be rewritten as 
\be
I(t_1)=\left(\frac{(1-\omega)^\omega}{2|C|}\right)^\frac{1}{1-\omega}\!\!\exp\left(\frac{2|C|\,t_0^{1-\omega}} {1-\omega}\right)
\Gamma\left(\frac{1}{1-\omega},\frac{2|C|\,t_1^{1-\omega}} {1-\omega}\right)\,,
\label{B-2}
\ee
where
\be
\Gamma(a,x)=\int_x^\infty u^{a-1}\e^{-u}\d u
\label{B-3}
\ee
is the incomplete gamma function.
 When $t_1\gg t_C$ we have $u(t_1)\gg1$. Thus, in this limit, the asymptotic expansion (see \cite{abramowitz65} p~263)
\be
\Gamma(a,x)\sim x^{a-1}\e^{-x}\left[1+\frac{a-1}{x}+\frac{(a-1)(a-2)}{x^2}+\cdots\right]\qquad (x\to\infty)
\label{B-4}
\ee
leads to:
\be
I(t_1)\sim\frac{t_1^\omega}{2|C|}\,\exp\!\left(-2|C|\,\frac{t_1^{1-\omega}-t_0^{1-\omega}} {1-\omega}\right)\!\left[1+\frac{\omega}{2|C|\,t_1^{1-\omega}}+\cdots\right]\,.
\label{B-5}
\ee
When $t_1\ll t_C$ then $u(t_1)\ll1$ and the series expansion (see \cite{abramowitz65} p~261--262)
\be
\Gamma(a,x)=\Gamma(a)-x^a\sum_{n=0}^\infty\frac{(-x)^n}{(a+n)n!}
\label{B-6}
\ee
can be used to write:
\be\fl
I(t_1)=\left(\frac{(1-\omega)^\omega}{2|C|}\right)^\frac{1}{1-\omega}\!\!\!\exp\!\left(\frac{2|C|\,t_0^{1-\omega}} {1-\omega}\right)\left[\Gamma\!\left(\frac{1}{\!1-\omega\!}\right)
-\left(\frac{2|C|}{(1-\omega)^\omega}\right)^\frac{1}{1-\omega}\!\!t_1+\cdots\right]\,.
\label{B-7}
\ee

\end{document}